\documentclass[1p,preprint,10pt]{elsarticle}
\usepackage{amsmath}
\usepackage{amssymb}
\usepackage[colorlinks=true,linkcolor=green,citecolor=red]{hyperref}
\usepackage{tikz}
\usepackage{braket}
\usepackage{multirow}
\usepackage{color}
\usepackage{listings}
\definecolor{light-gray}{gray}{0.95}
\definecolor{mygreen}{RGB}{0,153,0}
\journal{Computer Physics Communications}

\begin{document}

\begin{frontmatter}

\title{EDRIXS: An open source toolkit for simulating spectra of resonant inelastic x-ray scattering}

%% author list
\author[bnl]{Y.L. Wang\corref{cor1}} 
\ead{yilinwang@bnl.gov}
\author[bnl]{G. Fabbris}
\author[bnl]{M.P.M. Dean}
\author[bnl,rutgers]{G. Kotliar}
\cortext[cor1]{Corresponding author}

%% address list
\address[bnl]{Department of Condensed Matter Physics and Materials Science, Brookhaven National Laboratory, Upton, New York 11973, USA} 
\address[rutgers]{Department of Physics and Astronomy, Rutgers University, Piscataway, New Jersey 08856, USA}

\begin{abstract}
Resonant inelastic x-ray scattering (RIXS) has become a very powerful experimental technique to probe a broad range of intrinsic elementary excitations, for example, 
from low energy phonons and (bi-)magnons to high energy $d$-$d$, charge-transfer and plasmons excitations in strongly correlated electronic systems. 
Due to the complexity of the RIXS cross-section and strong core-hole effects, theoretical simulation of the experimental RIXS spectra is still a difficult task which
hampers the understanding of RIXS spectra and the development of the RIXS technique. 
In this paper, we present an open source toolkit (dubbed EDRIXS) to facilitate the simulations of RIXS spectra of strongly correlated materials based on exact diagonalization
(ED) of certain model Hamiltonians. The model Hamiltonian can be from a single atom, small cluster or Anderson impurity model, with model parameters from 
density functional theory plus Wannier90 or dynamical mean-field theory calculations.
The spectra of x-ray absorption spectroscopy (XAS) and RIXS are then calculated using Krylov subspace techniques. This toolkit contains highly efficient 
ED, XAS and RIXS solvers written in modern Fortran 90 language and a convenient Python library used to prepare inputs and set up calculations. 
We first give a short introduction to RIXS spectroscopy, and then we discuss the implementation details of this toolkit. Finally, we show three examples to demonstrate its usage.
\end{abstract}


\end{frontmatter}
\clearpage

\noindent\textbf{PROGRAM SUMMARY}

\noindent\textit{Program title:} EDRIXS

\noindent\textit{Catalogue identifier:} TO BE DONE

\noindent\textit{Program summary URL:} TO BE DONE

\noindent\textit{Program obtainable from:} TO BE DONE %%CPC Program Library, Queen’s University, Belfast, N. Ireland

\noindent\textit{Licensing provisions:} GNU General Public Licence 3.0

\noindent\textit{No. of lines in distributed program, including test data, etc.:} 28,365 lines

\noindent\textit{No. of bytes in distributed program, including test data, etc.:}  412,888 bytes

\noindent\textit{Distribution format:} tar.gz

\noindent\textit{Programming language:} Fortran 90 and Python3

\noindent\textit{Computer:} Desktop PC, laptop, high performance computing cluster

\noindent\textit{Operating system:} Unix, Linux, Mac OS X

\noindent\textit{Has the code been vectorised or parallelized?:} Yes, it is parallelized by MPI

\noindent\textit{RAM:} Heavily depends on the complexity of the problem

\noindent\textit{Classification:} 7.3

\noindent\textit{External routines/libraries used:} BLAS, LAPACK, Parallel ARPACK, numpy, scipy, sympy, matplotlib, sphinx, numpydoc

\noindent\textit{Nature of problem:} Simulating the spectra of resonant inelastic x-ray scattering for strongly correlated electronic systems. 

\noindent\textit{Solution method:}  Exact diagonalization of model Hamiltonians and Krylov subspace methods.

\noindent\textit{Running time:} Heavily depends on the complexity of the problem.

\clearpage

\section{Introduction\label{sec:intro}}
Directly measuring the elementary excitations in materials through experimental spectroscopy is a fundamental task in condensed matter physics.  
For example, angle-resolved photoemission spectroscopy (ARPES)~\cite{arpes_review:2003} is a very powerful technique to measure the single particle 
spectrum function $A(\vec{k},\omega)$ (bands) in both weakly and strongly correlated materials.
In recent years, resonant inelastic x-ray scattering (RIXS) has also emerged as a key probe of the 
elementary excitations in materials, especially for strongly correlated systems~\cite{Ament:2011}. RIXS is a \textit{photon-in photon-out} spectroscopy. 
The energy of incident photon is tuned at a specific resonant edge to excite a deep core electron to the valence shell and create a core-hole. 
After a very short time, the core-hole is refilled by a valence electron and photon is emitted. By measuring the change in energy, momentum and polarization 
of the scattered photon, one can obtain detailed information on the nature of the underlying excitations. 
Compared to other spectroscopies, RIXS has various unique features and advantages~\cite{Ament:2011}. First, it can measure a broad range of elementary excitations, 
for example, from low energy phonons and (bi-)magnons to high energy local $d$-$d$, charge transfer (CT) and plasmon excitations. 
Second, since x-rays carry appreciable momentum, the dispersion of low-energy excitations in solids can be probed.   
Third, there are fewer restrictions to sample surface quality or volume compared to probes such as ARPES and neutron scattering, 
which enables, for instance, studies of thin films and very small single crystals.
Finally, it is element and orbital specific, bulk sensitive and polarization dependent. 
In recent years, great progress in the technical development of RIXS has occurred due to high-brilliance synchrotron light sources and advanced photon detection instrumentation,
leading to its wide use to study elementary excitations in strongly correlated materials~\cite{Ament:2011, fabbris:2016, tomiyasu:2017, kim:2012, wray:2015}. 

Given the RIXS sensitivity to a wide range of elementary excitations, theoretical simulation of the RIXS cross-section is often very relevant in the interpretation
of the experimental RIXS spectra and its underlying physics. 
The RIXS cross-section is described by the Kramers-Heisenberg formula, which is a result of second-order perturbation theory. The RIXS cross-section can be
interpreted as a four-particle correlation function involving both valence and core electrons~\cite{shvaika:2005,igarashi:2013} and can be in principle determined by 
some first-principles numerical methods. Nowadays, density functional theory plus dynamical mean-field theory (DFT+DMFT)~\cite{georges:1996,kotliar:2006} with 
continuous-time quantum Monte-Carlo (CTQMC)~\cite{ctqmc:2011} as the impurity solver is the most powerful and accurate method for first-principle calculations of 
physical properites, such as the single particle spectrum function, of strongly correlated materials, so it is a desirable method to compute
RIXS spectra. Unfortunately, it is still a very challenging mission to determine the RIXS cross-section by this method with the current available numerical techniques 
due to the complexity of this correlation function and the strong unknown core-hole potential.
The difficulty in performing theoretical RIXS simulations hampers the progress of the understanding of RIXS spectra and the development of the RIXS technique.  
Nowadays, most of RIXS simulations are based on the exact diagonalization (ED) of Hamiltonians of small size models, that capture sort of the properites of the real materials, 
such as single atom, small cluster and Anderson impurity model (AIM)~\cite{de2008core}. Currently, the most widely used RIXS simulation codes based on ED algorithms include CTM4RIXS~\cite{CTM4XAS:2010} 
and Quanty~\cite{quanty:2016}. 
The former is based on an old atomic code called Cowan's code~\cite{cowan:1981} with very limited functionalities to perform XAS and RIXS calculations. 
The latter is based on a lightweight programming language called Lua~\cite{Lua} and has much richer functionalities. 
However, the core of the Quanty code is not open source, so it cannot be freely modified by users to implement functionalities that are not available. 
Many RIXS simulations based on ED~\cite{tsutsui:2003,ishii:2005,okada:2006,vernay:2008,chen:2010,forte:2011,kourtis:2012,uldry:2012,jia:2012,wohlfeld:2013,monney:2013,jia:2014,javier:2015,jia:2016,tsutsui:2016,green:2016,yuan:2017,kim:2017,paramekanti:2018,hariki:2018,umesh:2018,schlappa:2018} 
or DMRG~\cite{umesh:2018,Nocera2018,tohyama:2015} have been also reported. 
In Refs.~\cite{jia:2012,jia:2014,jia:2016}, 
the ED code is based on the parallel ARPACK library. In Refs.~\cite{hariki:2018}, a powerful ED solver based on quantum chemistry algorithm is used to diagonalize 
the AIM from a converged DFT+DMFT calculations with many bath sites and continuous charge excitations have been found in high-valence transition-metal oxides 
in their simulated RIXS spectra. However, currently these codes are not open source.

In this work, we present the EDRIXS open source toolkit that is aimed at facilitating RIXS simulations based on ED algorithms. 
EDRIXS is designed as a post-processing tool for the open source COMSCOPE project~\cite{comscope,sangkook:2018}.
It implements parallel ED solvers with very high efficiency based on the standard Lanczos algorithms and parallel version of ARPACK library. 
The XAS and RIXS spectra are then calculated parallelly using Krylov subspace algorithms. The core components of this toolkit are implemented using the modern Fortran 90 language 
and the parallelism is achieved by MPI. The code has a layered structure, thus new functionalities can be easily implemented by users who want to perform 
other types of x-ray scattering simulations. An application programming interface (API) is implemented by the popular Python3 language to prepare inputs and set up calculations. 
A pure Python ED solver is also implemented in the API for small size problem.
The rest of this paper is organized as follows: In section~\ref{sec:theory}, we introduce the basic theory and methods of RIXS simulations. In section~\ref{sec:impl}, 
we explain the implementation of the code in detail. In section~\ref{sec:install}, we show how to install and use this code. In section~\ref{sec:examples}, we show three 
examples to demonstrate the usage of this code. In section~\ref{sec:future}, we discuss the plans of future development of EDRIXS.

\section{Basic theory and methods of RIXS simulation\label{sec:theory}}
\subsection{Geometry of RIXS experiment}
\begin{figure}[!ht]
\includegraphics[width=0.95\textwidth]{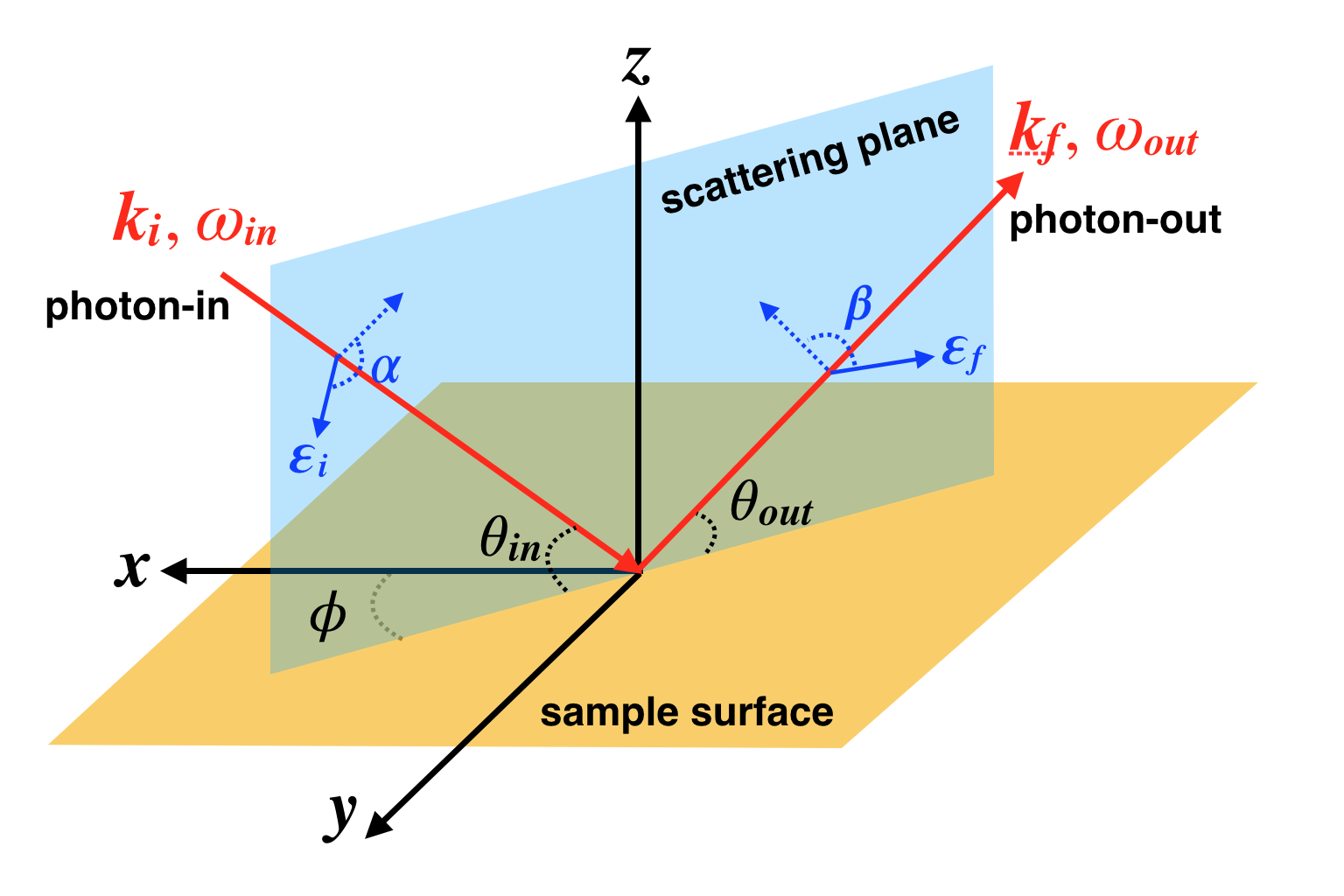}
\caption{The illustration of the geometry of RIXS experiment.}
\label{fig:geometry}
\end{figure}

A typical geometry of RIXS experiment is illustrated in Fig.~\ref{fig:geometry}. 
$\vec{k}_{i}$ ($\vec{\epsilon}_{i}$) and $\vec{k}_{f}$ ($\vec{\epsilon}_{f}$) are the wave (polarization) vectors of the incident and scattered x-ray, respectively.
$\omega_{\text{in}}$ and $\omega_{\text{out}}$ are the energy of incident and scattered x-ray, respectively.
$\alpha$ and $\beta$ are the angles between the polarization vectors and the scattering plane.
We call it $\pi$-polarization when the polarization vector is parallel to the scattering plane ($\alpha, \beta = 0$), and $\sigma$-polarization when the polarization vector is 
perpendicular to the scattering plane ($\alpha, \beta = \pi/2$).
$\theta_{\text{in}}$ and $\theta_{\text{out}}$ are the incident and scattered angles, respectively. $\phi$ is the azimuthal angle defined with respect to $x$-axis.

\subsection{XAS and RIXS cross-section}
The XAS cross-section is described by,
\begin{eqnarray}
I_{\text{XAS}}(\omega_{\text{in}}, \vec{\epsilon}_{i}) = \sum_{i} \frac{1}{Z} e^{-\frac{E_i}{K_BT}}\sum_{n}\left|\Braket{n|\hat{\mathcal{D}}_{i}|i}\right|^2 \frac{\Gamma_{c}/\pi}{(\omega_{\text{in}} - E_{n} + E_{i})^2 + \Gamma_{c}^{2}},
\end{eqnarray}
and the RIXS cross-section is described by the Kramers-Heisenberg formula~\cite{Ament:2011},
\begin{eqnarray}
I_{\text{RIXS}}(\omega_{\text{in}},\omega,\vec{k}_i, \vec{k}_f, \vec{\epsilon}_i, \vec{\epsilon}_f) &=& \sum_{i} \frac{1}{Z} e^{-\frac{E_i}{K_BT}}\sum_{f}\left|\Braket{f|\hat{\mathcal{D}}^{\dagger}_{f}\frac{1}{\omega_{\text{in}}-\hat{H}_{n}+E_i+i\Gamma_c}\hat{\mathcal{D}}_{i}|i}\right|^2 \nonumber \\
           & \times &\frac{\Gamma /\pi}{(\omega-E_f+E_i)^2+\Gamma^2}.
\end{eqnarray}
where, $\Ket{i}$ are the ground states of the Hamiltonian $\hat{H}_{i}$ describing the valence electrons. $T$ is temperature and $K_{B}$ is the Boltzmann factor. 
$Z=\sum_{i}e^{-\frac{E_i}{K_{B}T}}$ is the partition function. 
$\Ket{f}$ are the excited eigenstates of $\hat{H}_{i}$.
$\Ket{n}$ are the eigenstates of intermediate Hamiltonian $\hat{H}_{n}$ which includes a deep core-hole and one more electron in the valence shell 
after the x-ray absorption process. 
$\omega_{\text{in}}$ is the energy of the incident x-ray which is tuned to be at a specific resonant edge. $\omega$ is the x-ray energy loss, i.e. the energy difference between
the incident and scattered x-ray. $\Gamma_{c}$ and $\Gamma$ are the lifetime broadening of the intermediate (with a core hole) and final states, respectively.

$\hat{\mathcal{D}}_i$ and $\hat{\mathcal{D}}^{\dagger}_{f}$ are the transition operators for x-ray absorption and emission process, respectively. 
They can be generally written as,
\begin{eqnarray}
\hat{\mathcal{D}}_i &=& \sum_{a}P_{a}^{i}\hat{T}_{a}^{i}, \\
\hat{\mathcal{D}}_{f}^{\dagger} &=& \sum_{a}P_{a}^{f\star}\hat{T}_{a}^{f\dagger}.
\end{eqnarray}
where, $P_{a}^{i}$ and $P_{a}^{f}$ involve the polarization and wave vector of x-ray. For dipolar transition, they are: 
$\epsilon_{x}^{i}, \epsilon_{y}^{i}, \epsilon_{z}^{i}$ and $\epsilon_{x}^{f}, \epsilon_{y}^{f}, \epsilon_{z}^{f}$. 
$\hat{T}_{a}^{i}$ and $\hat{T}_{a}^{f}$ are components of transition operators. For dipolar transition, they are: 
\begin{eqnarray}
\hat{T}_{x}^{i} &=& \sum_{R}e^{i\vec{k}_{i}\cdot\vec{R}}\hat{x}_{R},\;\; \hat{T}_{y}^{i} = \sum_{R}e^{i\vec{k}_{i}\cdot\vec{R}}\hat{y}_{R}, \;\; \hat{T}_{z}^{i} = \sum_{R}e^{i\vec{k}_{i}\cdot\vec{R}}\hat{z}_{R}, \\
\hat{T}_{x}^{f} &=& \sum_{R}e^{i\vec{k}_{f}\cdot\vec{R}}\hat{x}_{R},\;\; \hat{T}_{y}^{f} = \sum_{R}e^{i\vec{k}_{f}\cdot\vec{R}}\hat{y}_{R}, \;\; \hat{T}_{z}^{f} = \sum_{R}e^{i\vec{k}_{f}\cdot\vec{R}}\hat{z}_{R},
\end{eqnarray}
where, $\vec{R}$ is site-index, and $\hat{x}_R$, $\hat{y}_{R}$ and $\hat{z}_{R}$ are position operators of electrons bound to site-$R$. 
These are measured with respect to the center of site-$R$. 

\subsection{Model Hamiltonian}
$\hat{H}_i$ is a model Hamiltonian for valence electrons from single atom, cluster or Anderson impurity models. It can be generally witten as,
\begin{equation}
\hat{H}_{i} = \sum_{\alpha,\beta} t_{\alpha,\beta} \hat{f}^{\dagger}_{\alpha} \hat{f}_{\beta} + \sum_{\alpha,\beta,\gamma,\delta} U_{\alpha,\beta,\gamma,\delta} \hat{f}^{\dagger}_{\alpha}\hat{f}^{\dagger}_{\beta}\hat{f}_{\gamma}\hat{f}_{\delta},
\end{equation} 
where, $t_{\alpha,\beta}$ usually contain crystal field (CF) splitting terms, spin-orbit coupling (SOC) terms, hopping terms between different sites. 
The CF and hopping terms can be obtained from a first-principle DFT+Wannier90 calculations. SOC can be obtained from DFT calculations or single atomic calculation, 
for example, from Cowan's atomic code~\cite{cowan:1981}. $U_{\alpha,\beta,\gamma,\delta}$ are the rank-4 tensors of the on-site Coulomb interaction and are parameterized 
by Slater integrals, which can be obtained from an atomic calculations. 

$\hat{H}_{n}$ is the intermediate Hamiltonian with a core-hole, which can be written as, 
\begin{equation}
\hat{H}_{n} = \hat{H}_{i} + \hat{V}_{\text{core-hole}} + \hat{H}_{\text{core}},
\end{equation}
where, the core-hole potential $\hat{V}_{\text{core-hole}}$ is simulated at atomic level by adding Coulomb interaction between valence electrons and core-hole.  
$\hat{H}_{\text{core}}$ is the Hamiltonian of the core electrons.

To simulate RIXS spectra based on DFT+DMFT calculations, we first perform a DFT+DMFT calculation to get converged hybridization function $\Delta_{\alpha}(i\omega_{n})$ and 
then construct an Anderson impurity model $\hat{H}_{i}$ from it, which can be written as,
\begin{equation}
\hat{H}_{i} = \hat{H}_{\text{imp}} + \sum_{\alpha,l}E_{\alpha,l}\hat{c}^{\dagger}_{\alpha,l}\hat{c}_{\alpha,l} + \sum_{\alpha,l}V^{l}_{\alpha}\hat{f}^{\dagger}_{\alpha}\hat{c}_{\alpha,l} + h.c. ,
\end{equation}
where, $\hat{H}_{\text{imp}}$ is the Hamiltonian for the impurity site, which is taken from the atomic problem in DFT+DMFT calculation.
$E_{\alpha,l}$ is the energy level of the $l$-th bath site with orbital $\alpha$, $V^{l}_{\alpha}$ are the hybridization strength between localized impurity 
electrons ($\hat{f}^{\alpha}$) and conducted bath electrons ($\hat{c}_{\alpha,l}$). $E_{\alpha,l}$ and $V_{\alpha}^{l}$ are obtained by fitting the hybridization function,
\begin{equation}
\Delta_{\alpha}(i\omega_{n}) = \sum_{l=1}^{N} \frac{\left|V_{\alpha}^{l}\right|^2}{i\omega_{n}-E_{\alpha,l}}.
\end{equation}
The core-hole potential is only added to the impurity site in $\hat{H}_{n}$.

\subsection{RIXS simulation algorithms}
\begin{figure}[!ht]
\includegraphics[width=0.95\textwidth]{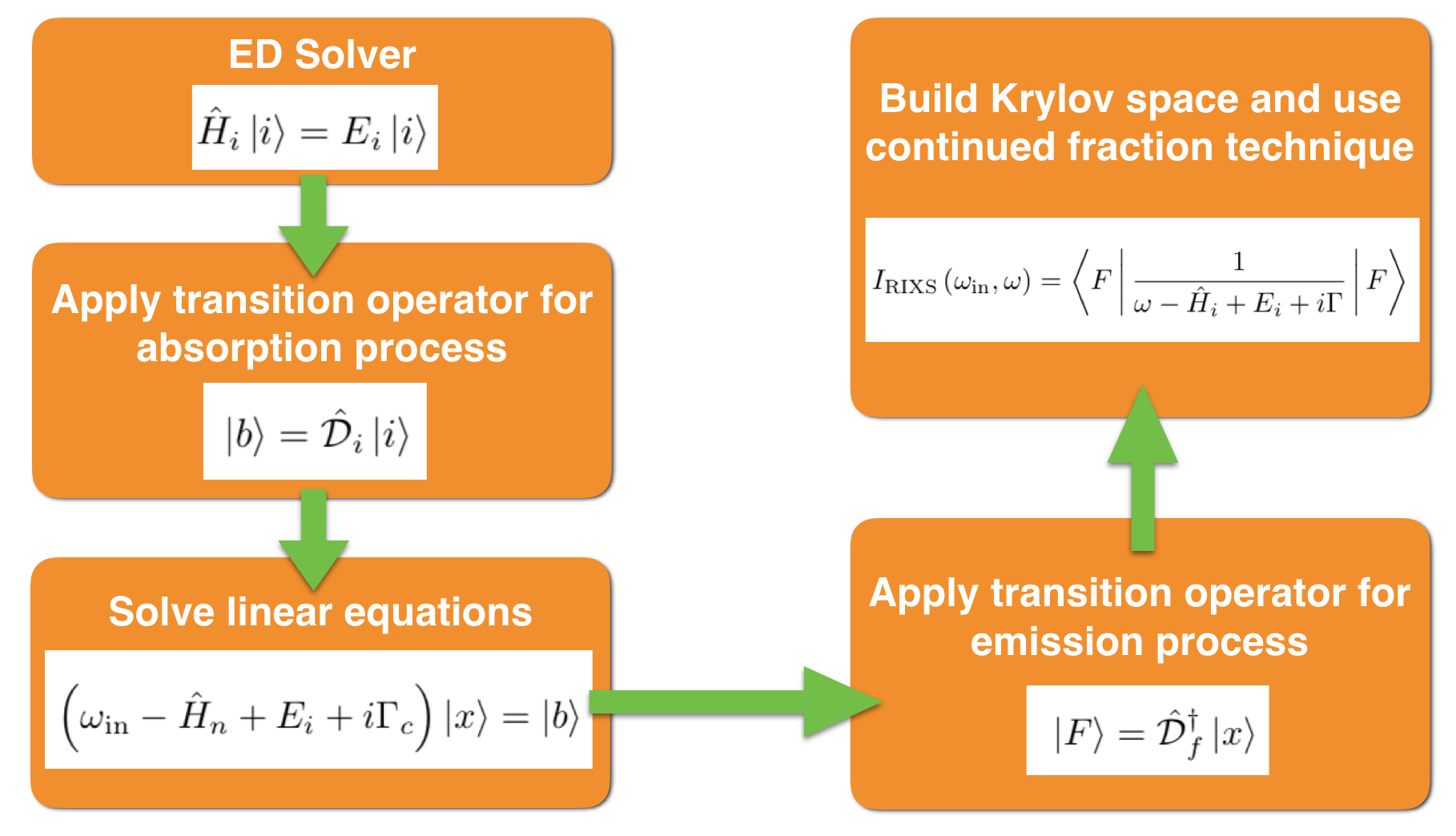}
\caption{The flow diagram of algorithm to calculate RIXS cross-section.}
\label{fig:flow}
\end{figure}

We use similar RIXS simulation algorithms as that used in Ref.~\cite{jia:2012,hariki:2018}.
The flow diagram of the algorithm to calculate the RIXS cross-section is illustrated in Fig.~\ref{fig:flow} and briefly described as follows:
\begin{itemize}
\item Use an ED solver to diagonalize $\hat{H}_i$ to get the ground states or a few low-energy excited states $\Ket{i} (E_i)$.
Both standard Lanczos algorithm~\cite{Lanczos:1950} without re-orthogonalization and implicitly restarted Arnoldi algorithm~\cite{Arnoldi:1951,Sorensen1997}
are used in ED solvers. 

\item Apply the transition operator for the x-ray absorption process on $\Ket{i}$ to get $\Ket{b}=\hat{\mathcal{D}}_i\Ket{i}$. 

\item Solve the following large-scale sparse symmetric linear equations,
\begin{equation}
\left(\omega_{\text{in}} - \hat{H}_{n} + E_i + i\Gamma_c\right)\Ket{x} = \Ket{b},
\end{equation}
by minimum residual (MINRES) method~\cite{minres}.

\item Apply the transition operator for the x-ray emission process on $\Ket{x}$ to get $\Ket{F}=\hat{\mathcal{D}}^{\dagger}_{f}\Ket{x}$. 

\item Build Krylov space of $\hat{H}_{i}$ and use continued fraction technique to get the RIXS spectra,
\begin{equation}
I_{\text{RIXS}}(\omega_{\text{in}}, \omega) = \Braket{F|\frac{1}{\omega -\hat{H}_{i} + E_i+i\Gamma}|F}.
\end{equation}

\end{itemize}

\begin{figure}[!ht]
\includegraphics[width=0.95\textwidth]{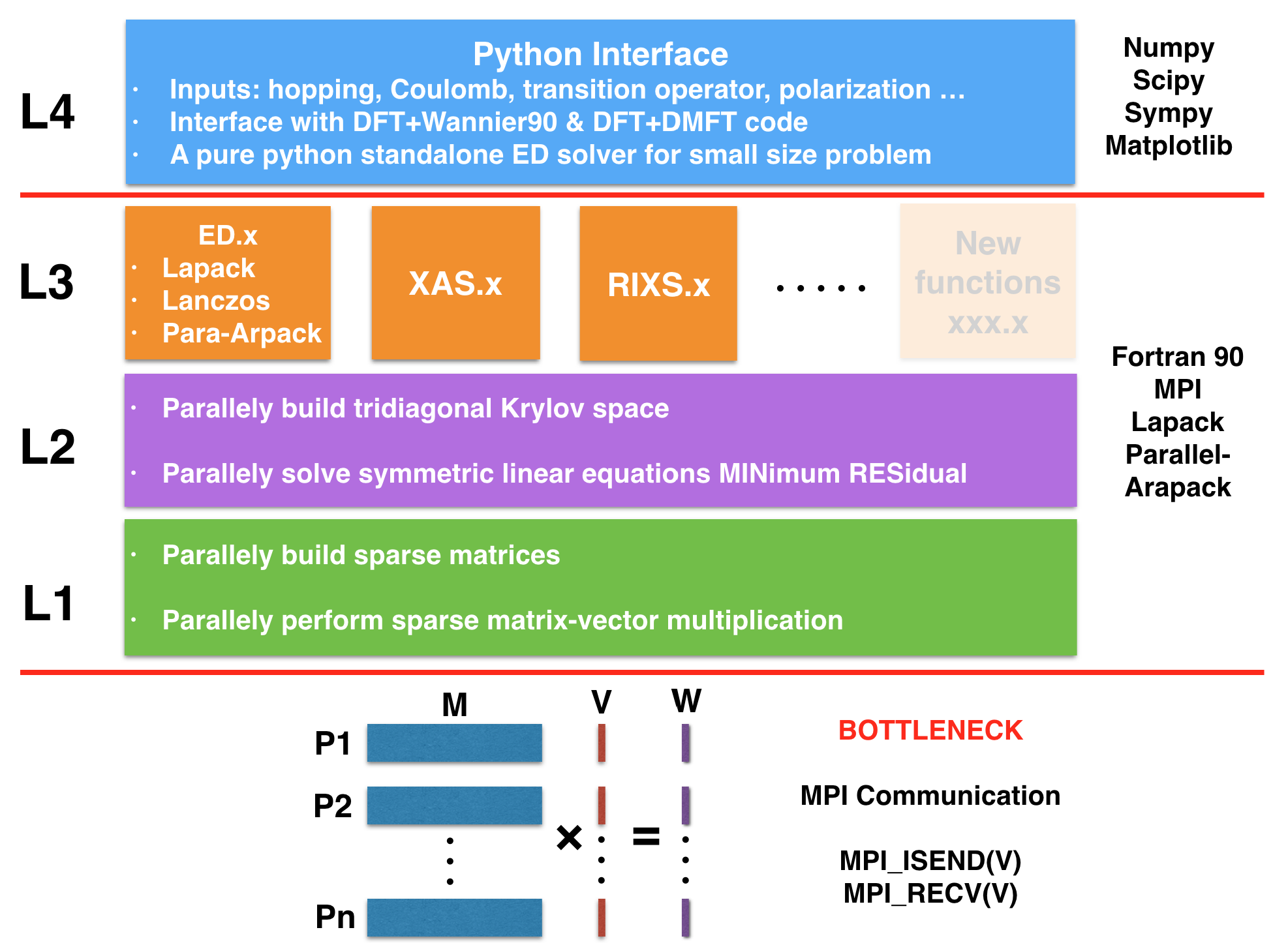}
\caption{The structure of EDRIXS code.}
\label{fig:struct}
\end{figure}

\section{Implementations and optimizations\label{sec:impl}}
In this section, we describe the details of implementation and optimization of EDRIXS.

\subsection{Development platform}
The key components of the EDRIXS code are developed with the modern Fortran 90 language. Intel's \textit{ifort} and GNU's \textit{gfortran} compilers have been tested.
The code is parallelized by MPI. An interface that is used for pre- and post-processing is written by Python3 language. Git is used as the version
control system. 

\subsection{Fock basis}
We implement EDRIXS code based on the second-quantization language. We first define a single particle basis $\alpha$, then define Fock basis $\Ket{I}$ with respect to it.
In the code, all the orbitals of valence electrons are gathered together and put in front of all the orbitals of core electrons 
(i.e. indices of valence orbitals are always smaller than those of core orbitals). 
A Fock state can be represented by a binary number with digital 1 or 0, where 1 means that the orbital is occupied and 0 means that the orbital is empty. For example,
\begin{equation}\label{eqn:binary}
\overset{1-6}{\overbrace{{\color{red}110100}}}\;\overset{7-12}{\overbrace{{\color{blue}111011}}}
\end{equation}
where, the red parts with orbital indices from 1 to 6 are for valence electrons and the blue parts with orbital indices from 7 to 12 are for core electrons.
In this example, valence orbitals 1, 2 and 4 are occupied, and one core orbital 10 is empty. The above binary format is used for the ED solver implemented in Python API. 
For \textit{ed.x}, \textit{xas.x} and \textit{rixs.x}, we need to input the Fock basis with their corresponding decimal formats, but only for the parts of valence orbitals. 
The core orbitals will be explicitly considered within these three programs. For the above example (Eqn.~\ref{eqn:binary}), the corresponding decimal number is computed as 
following,
\begin{equation}
\textcolor{red}{1}\times 2^{1-1} + \textcolor{red}{1}\times 2^{2-1} + \textcolor{red}{0}\times 2^{3-1} + \textcolor{red}{1}\times 2^{4-1} + \textcolor{red}{0}\times 2^{5-1} + \textcolor{red}{0}\times 2^{6-1} = 11.
\end{equation}

\subsection{Format of sparse matrices}
In second-quantization form, a general operator can be written as,
\begin{equation}
\hat{O} = \sum_{\alpha,\beta} t_{\alpha,\beta} \hat{f}^{\dagger}_{\alpha}\hat{f}_{\beta} + \sum_{\alpha,\beta,\gamma,\delta}U_{\alpha,\beta,\gamma,\delta}\hat{f}^{\dagger}_{\alpha}\hat{f}^{\dagger}_{\beta}\hat{f}_{\gamma}\hat{f}_{\delta},
\end{equation}
where, we call $t_{\alpha,\beta}$ hopping terms and $U_{\alpha,\beta,\gamma,\delta}$ rank-4 interaction terms.
The matrix elements of $\hat{O}$ in the Fock basis $\Ket{I}$ are $\Braket{I^{\prime}|\hat{O}|I}$. 
These matrices are large-scale sparse matrices and the number of nonzero elements is proportional to its dimension $n$, so we store them in compressed sparse row (CSR) format
to save memory. To speed up the building of operators, only nonzeros of $t_{\alpha,\beta}$ and $U_{\alpha,\beta,\gamma,\delta}$ are input.

\subsection{Structure of the code}
The structure of the EDRIXS code can be divided into four layers, which is illustrated in Fig.~\ref{fig:struct}.
The key components of EDRIXS code involve two basic operations: building matrices form of operators in Fock basis and performing sparse matrix-vector multiplication (MVM), 
which are the most time-consuming parts, so we implement subroutines to parallelly do these tasks in the first layer. 
The rows of a sparse matrix $M$ and elements of a vector $V$ are equally distributed among all the processors. Each processor only needs to build parts of the rows of
the operator and do parts of the MVM, and all the processors can work parallelly. The only MPI communication needed for MVM is to send and receive the vector $V$, which 
is a bottleneck of the code. The speed of this MPI communication heavily depends on the networking of the cluster. 
Based on the first layer, we implement subroutines to parallelly build tridiagonal Krylov space and solve symmetric linear equations using MINRES methods in the 
second layer. The key operations of these two subroutines are MVM.

Above the first and second layers, we implement the main functionalities: \textit{ed.x}, \textit{xas.x} and \textit{rixs.x} to do RIXS simulations in the third layer. 
Three types of ED solvers are implemented. 
The first one is to use the LAPACK's subroutine ZHEEV to diagonalize a Hamiltonian to get all the eigenstates, which is useful for small size ($n<10,000$) problem. 
The second one is to use standard Lanczos algorithm without re-orthogonalization to get only one ground state, which is not very accurate due to the orthogonality problem, 
especially for degenerate ground states. However, it runs very fast and is suitable for roughly searching the ground state of a Hamiltonian in many different subspaces. 
The third one is to use the parallel version of the ARPACK library to get few lowest eigenstates, which is very accurate because it uses implicitly restarted Arnoldi 
algorithm to deal with the orthogonalization issue, but its efficiency is lower than the second one. We will use this solver in real RIXS simulations.
Based on the first and second layers, we can easily extend new functionalities to do more calculations, for example, calculating the expected value of an operator 
in the ground state and other correlation functions.

In the fourth layer, a Python interface is implemented to provide some application programming interfaces (APIs) for preparing inputs for \textit{ed.x}, \textit{xas.x} and 
\textit{rixs.x} and setting up calculations. Besides, a pure python ED solver based on second-quantization is also implemented for small size problem, so the RIXS simulations can be done easily with pure Python code. Currently, we have not implemented the interface with the Cowan's code for calculating Slater integrals. One should 
manually run Cowan's code and read Slater integrals. We refer users to its official webpage~\cite{cowan_web} for more information.

\section{Installation and usage\label{sec:install}}
In this section, we will show how to install and use EDRIXS.

\subsection{Get EDRIXS}
EDRIXS is an open source code and is released under the GNU General Public Licence 3.0 (GPL). The readers who are interested in it can write a letter to the authors 
to request a copy of the newest version of EDRIXS source code, or they can download it directly from the public code repository:
\begin{verbatim}
    $ git clone https://github.com/shenmidelin/edrixs.git.
\end{verbatim}
where \$ is the command line prompt. One can use the command 
\begin{verbatim}
    $ git pull 
\end{verbatim}
to get the latest updates of the code.

\subsection{Build EDRIXS}
To build the source code and documentation of EDRIXS, some tools and external libraries are required, which are listed in Table~\ref{table:libs}.
Most of the libraries can be easily installed using \textit{Anaconda}~\cite{anaconda}.  
There may be problems when using \textit{gfortran} with \textit{MKL}, so we strongly recommend \textit{gfortran}+\textit{OpenBLAS}.
Note that the \textit{arpack-ng} library should be compiled with the same Fortran compiler and \textit{BLAS/LAPACK} libraries.
\begin{table}[!ht]
\caption{Requirement of tools and external libraries for building EDRIXS code. \label{table:libs}}
\centering
\begin{tabular}{l|l}
\hline
Tools and Libraries       & Recommendation  \\ \hline
Fortran compiler          & Intel's ifort or GNU's gfortran  \\ 
MPI                       & OpenMPI or MPICH \\
BLAS/LAPACK               & MKL or OpenBLAS~\cite{openblas} \\ 
Parallel ARPACK           & arpack-ng~\cite{arpackng}, v3.6.2 or higher \\
Python3                   & v3.6 or higher \\ 
Numpy                     & v1.15 or higher \\ 
Scipy                     & v1.1 or higher \\ 
Sympy                     & v1.3 or higher \\ 
Matplotlib                & v3.0 or higher  \\ 
Sphinx                    & v1.7 or higher  \\ 
Numpydoc                  & v0.8 or higher \\ \hline
\end{tabular}
\end{table}

We assume that one downloads the source code into the directory \textit{EDRIXS\_DIR}.
One then go to the \textit{edrixs/src/fortran} directory and edit the \textit{make.sys} file to configure the compiling settings.
\begin{verbatim}
    $ cd edrixs/src/fortran
    $ cp make.sys.ifort make.sys  (or cp make.sys.gfortran make.sys)
    $ vim make.sys
\end{verbatim}
where, in the \textit{make.sys} file, one needs to set F90 as
\begin{verbatim}
    F90 = mpif90 
\end{verbatim}
and set LIBS, for example,
\begin{verbatim}
    LIBS = -L${MKLROOT}/lib/intel64 -lmkl_core -lmkl_sequential -lmkl_rt \\
           -L${ARPACK_DIR}/lib/ -lparpack -larpack
\end{verbatim}
where, we assume the \textit{ARPACK} library is installed in \textit{ARPACK\_DIR}. Then, we type
\begin{verbatim}
    $ make
    $ make install
\end{verbatim}
to compile and install the executable files \textit{ed.x}, \textit{xas.x} and \textit{rixs.x} into \textit{bin} directory. 
Then, one should add the following two lines in \textit{.bashrc} or \textit{.bash\_profile} file,
\begin{verbatim}
    export PATH=${EDRIXS_DIR}/edrixs/bin:$PATH
    export PYTHONPATH=${EDRIXS_DIR}/edrixs/src/python:$PYTHONPATH
\end{verbatim}
To compile the documentation of the Python APIs, one should
\begin{verbatim}
    $ cd edrixs/docs
    $ mkdir build
    $ sphinx-build -b html source build 
    $ make html
\end{verbatim}
and open the file
\begin{verbatim}
    ${EDRIXS_DIR}/edrixs/docs/build/index.html
\end{verbatim}
in a browser to read the documentation of the Python APIs. 

\subsection{Use EDRIXS}
The typical procedures of the usage of EDRIXS are as follows, 
\begin{itemize}
\item Use Python APIs to set up the parameters of Hamiltonian, transition operators, Fock basis.
\item Run \textit{ed.x}, \textit{xas.x} and \textit{rixs.x} in order. These three Fortran executable files can be launched manually, for example,
\begin{verbatim}
    mpirun -np 16 ed.x > log.dat
\end{verbatim}
or through the Python \textit{subprocess} call, for example,
\begin{verbatim}
    import subprocess
    subprocess.check_call([``mpirun", ``-np", ``16", ``ed.x"]) 
\end{verbatim}
\item Use Python APIs to post-process the results and plot XAS or RIXS spectra.
\end{itemize}
The required input and output files for \textit{ed.x}, \textit{xas.x} and \textit{rixs.x} are listed in Table~\ref{table:ed}, Table~\ref{table:xas} and Table~\ref{table:rixs}, respectively. 
More details about the input and output files are described in file \textit{user\_guide.pdf} in \textit{docs} directory. 

\begin{table}[!ht]
\caption{Input and output files for \textit{ed.x}. \label{table:ed}}
\centering
\begin{tabular}{l|l|l}
\hline
                                             & File Name     & Description              \\ \hline
\multicolumn{1}{c|}{\multirow{4}{*}{Inputs}} & config.in     & set up control parameters \\  
\multicolumn{1}{c|}{}                        & fock\_i.in    & Fock basis $\Ket{I}$ for $\hat{H}_{i}$    \\ 
\multicolumn{1}{c|}{}                        & hopping\_i.in & $t_{\alpha,\beta}$ terms in $\hat{H}_{i}$ \\ 
\multicolumn{1}{c|}{}                        & coulomb\_i.in &  $U_{\alpha,\beta,\gamma,\delta}$ terms in $\hat{H}_{i}$\\ \hline
\multirow{4}{*}{Outputs}                     & standard outputs & log file \\ 
                                             & eigvals.dat   & eigenvalues $E_i$ \\ 
                                             & eigvec.$i$ &  the $i$-th ground state $\Ket{\Gamma_i}$ \\ 
                                             & denmat.dat &  density matrix: $\Braket{\Gamma_{i}|\hat{f}^{\dagger}_{\alpha}\hat{f}_{\beta}|\Gamma_{i}}$\\ \hline
\end{tabular}
\end{table}

\begin{table}[!ht]
\caption{Input and output files for \textit{xas.x}. \label{table:xas}}
\centering
\begin{tabular}{l|l|l}
\hline
                                             & File Name     & Description              \\ \hline
\multicolumn{1}{c|}{\multirow{7}{*}{Inputs}} & config.in     & set up control parameters \\ 
\multicolumn{1}{c|}{}                        & fock\_i.in    & Fock basis $\Ket{I}$ for $\hat{H}_{i}$     \\ 
\multicolumn{1}{c|}{}                        & fock\_n.in    & Fock basis $\Ket{I}$ for $\hat{H}_{n}$     \\ 
\multicolumn{1}{c|}{}                        & hopping\_n.in & $t_{\alpha,\beta}$ terms in $\hat{H}_{n}$ \\ 
\multicolumn{1}{c|}{}                        & coulomb\_n.in &  $U_{\alpha,\beta,\gamma,\delta}$ terms in $\hat{H}_{n}$\\ 
\multicolumn{1}{c|}{}                        & transop\_xas.in & transition operator $\hat{\mathcal{D}}_{i}$ \\ 
\multicolumn{1}{c|}{}                        & eigvec.$i$ &  the $i$-th ground state $\Ket{\Gamma_{i}}$ of $\hat{H}_{i}$\\ \hline
\multirow{2}{*}{Outputs}                     & standard outputs & log file \\ 
                                             & xas\_poles.$i$ &  XAS data for the $i$-th ground state \\ \hline
\end{tabular}
\end{table}

\begin{table}[!ht]
\caption{Input and output files for \textit{rixs.x}. \label{table:rixs}}
\centering
\begin{tabular}{l|l|l}
\hline
                                             & File Name     & Description              \\ \hline
\multicolumn{1}{c|}{\multirow{11}{*}{Inputs}} & config.in     & set up control parameters \\ 
\multicolumn{1}{c|}{}                        & fock\_i.in    & Fock basis $\Ket{I}$ for $\hat{H}_{i}$     \\ 
\multicolumn{1}{c|}{}                        & fock\_n.in    & Fock basis $\Ket{I}$ for $\hat{H}_{n}$     \\
\multicolumn{1}{c|}{}                        & fock\_f.in    & Fock basis $\Ket{I}$ for $\hat{H}_{i}$     \\
\multicolumn{1}{c|}{}                        & hopping\_i.in & $t_{\alpha,\beta}$ terms in $\hat{H}_{i}$ \\
\multicolumn{1}{c|}{}                        & hopping\_n.in & $t_{\alpha,\beta}$ terms in $\hat{H}_{n}$ \\
\multicolumn{1}{c|}{}                        & coulomb\_i.in &  $U_{\alpha,\beta,\gamma,\delta}$ terms in $\hat{H}_{i}$\\ 
\multicolumn{1}{c|}{}                        & coulomb\_n.in &  $U_{\alpha,\beta,\gamma,\delta}$ terms in $\hat{H}_{n}$\\ 
\multicolumn{1}{c|}{}                        & transop\_rixs\_i.in & transition operator $\hat{\mathcal{D}}_{i}$ \\ 
\multicolumn{1}{c|}{}                        & transop\_rixs\_f.in & transition operator $\hat{\mathcal{D}}_{f}^{\dagger}$ \\ 
\multicolumn{1}{c|}{}                        & eigvec.$i$ &  the $i$-th ground state $\Ket{\Gamma_{i}}$ of $\hat{H}_{i}$\\ \hline
\multirow{2}{*}{Outputs}                     & standard outputs & log file \\ 
                                             & rixs\_poles.$i$ &  RIXS data for the $i$-th ground state \\ \hline
\end{tabular}
\end{table}

\section{Examples\label{sec:examples}}
In this section, we show three examples to demonstrate the usage of EDRIXS. 

\subsection{Hello RIXS: multiplets $d$-$d$ excitations in a single atom case Ni ($d^{8}$)}
In this example, we use pure Python code to do ED calculation and calculate XAS and RIXS spectra step by step. The main purpose is to show how the Python APIs are used.
This example is a resonant x-ray scattering at Ni-$L_{2/3}$ edge ($2p_{1/2,3/2}\rightarrow 3d$ transition). Ni has a $d^{8}$ configuration with a tetragonal CF environment~\cite{fabbris:2016}.
The Python codes are shown in the Listing~\ref{code:ex1}.
\begin{lstlisting}[language=Python, caption=A Python example to calculate XAS and RIXS, label={code:ex1}]
#!/usr/bin/env python

import numpy as np
import matplotlib.pyplot as plt
from matplotlib.ticker import MultipleLocator, FormatStrFormatter
from soc               import atom_hsoc
from utils             import boltz_dist
from rixs_utils        import scattering_mat
from fock_basis        import get_fock_bin_by_N
from coulomb_utensor   import get_umat_slater, get_F0
from basis_transform   import cb_op, cb_op2, tmat_r2c
from photon_transition import get_trans_oper, dipole_polvec_rixs
from manybody_operator import two_fermion, four_fermion

font = {'family'  : 'Times New Roman' ,
         'weight' : 'medium' ,
         'size'   : '18'}
plt.rc('font',**font)

def ed():
    # 1-10: Ni-3d valence orbitals, 11-16: Ni-2p core orbitals 
    # Single particle basis: complex shperical Harmonics
    ndorb, nporb, ntot = 10, 6, 16
    emat_i = np.zeros((ntot,ntot), dtype=np.complex)
    emat_n = np.zeros((ntot,ntot), dtype=np.complex)

    # 4-index Coulomb interaction tensor, parameterized by 
    # Slater integrals, which are obtained from Cowan's code
    # Interaction among 3d valence electrons
    # averaged dd Coulomb interaction is set to be zero
    F2_d, F4_d   = 7.9521,  4.9387
    F0_d         = get_F0('d', F2_d, F4_d)

    # Core-hole potential: interaction between 3d and 2p
    # averaged dp Coulomb interaction is set to be zero
    G1_dp, G3_dp = 4.0509, 2.3037
    F0_dp, F2_dp = get_F0('dp', G1_dp, G3_dp), 7.33495

    umat_i = get_umat_slater('dp', F0_d, F2_d, F4_d,  # dd
                                   0, 0, 0, 0,        # dp
                                   0, 0)              # pp
    umat_n = get_umat_slater('dp', F0_d, F2_d, F4_d,           # dd
                                   F0_dp, F2_dp, G1_dp, G3_dp, # dp 
                                   0, 0)                       # pp

    # Atomic spin-orbit coupling
    zeta_d, zeta_p = 0.083, 11.24
    emat_i[0:ndorb, 0:ndorb]       += atom_hsoc('d', zeta_d)
    emat_n[0:ndorb, 0:ndorb]       += atom_hsoc('d', zeta_d)
    emat_n[ndorb:ntot, ndorb:ntot] += atom_hsoc('p', zeta_p)

    # Tetragonal crystal field splitting terms,
    # which are first defined in the real cubic Harmonics basis,
    # and then transformed to complex shperical Harmonics basis.
    dt, ds, dq = 0.011428, 0.035714, 0.13
    tmp = np.zeros((5,5),dtype=np.complex)
    tmp[0,0] =  6*dq - 2*ds - 6*dt   # d3z2-r2
    tmp[1,1] = -4*dq - 1*ds + 4*dt   # dzx
    tmp[2,2] = -4*dq - 1*ds + 4*dt   # dzy
    tmp[3,3] =  6*dq + 2*ds - 1*dt   # dx2-y2
    tmp[4,4] = -4*dq + 2*ds - 1*dt   # dxy
    tmp[:,:] = cb_op(tmp, tmat_r2c('d'))
    emat_i[0:ndorb:2, 0:ndorb:2] += tmp
    emat_i[1:ndorb:2, 1:ndorb:2] += tmp
    emat_n[0:ndorb:2, 0:ndorb:2] += tmp
    emat_n[1:ndorb:2, 1:ndorb:2] += tmp

    # Build Fock basis in its binary form 
    basis_i = get_fock_bin_by_N(ndorb, 8, nporb, nporb)
    basis_n = get_fock_bin_by_N(ndorb, 9, nporb, nporb-1)
    ncfg_i, ncfg_n = len(basis_i), len(basis_n)

    # Build many-body Hamiltonian in Fock basis
    hmat_i = np.zeros((ncfg_i,ncfg_i), dtype=np.complex)
    hmat_n = np.zeros((ncfg_n,ncfg_n), dtype=np.complex)
    hmat_i[:,:] += two_fermion(emat_i,  basis_i, basis_i)
    hmat_i[:,:] += four_fermion(umat_i, basis_i)
    hmat_n[:,:] += two_fermion(emat_n,   basis_n, basis_n)
    hmat_n[:,:] += four_fermion(umat_n, basis_n)

    # Do exact-diagonalization to get eigenvalues and eigenvectors
    eval_i, evec_i = np.linalg.eigh(hmat_i)
    eval_n, evec_n = np.linalg.eigh(hmat_n)

    # Build dipolar transition operators
    dipole = np.zeros((3, ntot, ntot), dtype=np.complex)
    T_abs = np.zeros((3, ncfg_n, ncfg_i), dtype=np.complex)
    T_emi = np.zeros((3, ncfg_i, ncfg_n), dtype=np.complex)
    tmp = get_trans_oper('dp')
    for i in range(3):
        dipole[i, 0:ndorb, ndorb:ntot] = tmp[i]
        # First, in the Fock basis
        T_abs[i] = two_fermion(dipole[i], basis_n, basis_i)
        # Then, transfrom to the eigenvector basis
        T_abs[i] = cb_op2(T_abs[i], evec_n, evec_i)
        T_emi[i] = np.conj(np.transpose(T_abs[i]))

    return eval_i, eval_n, T_abs, T_emi

def xas(eval_i, eval_n, T_abs): 
    gs = list(range(0, 3)) # three lowest eigenstates are used
    prob = boltz_dist([eval_i[i] for i in gs], 300) 
    off = 857.4  # offset of the energy of the incident x-ray 
    Gam_c = 0.2  # core-hole life-time broadening
    pol = np.array([1.0, 1.0, 1.0])/np.sqrt(3.0) # isotropic
    omega = np.linspace(-10, 20, 1000) 
    xas = np.zeros(len(omega), dtype=np.float)
    # Calculate XAS spectrum
    for i, om in enumerate(omega):
        for j in gs:
            F_mag = (T_abs[0,:,j]*pol[0] + 
                     T_abs[1,:,j]*pol[1] + 
                     T_abs[2,:,j]*pol[2] )

            xas[i] += prob[j]*np.sum(np.abs(F_mag)**2 * 
                      Gam_c/np.pi/((om-(eval_n[:]-eval_i[j]))**2 
                      + Gam_c**2))

    # plot XAS
    plt.figure()
    ax = plt.subplot(1,1,1)
    plt.ylim([0,0.6])
    plt.plot(omega+off, xas, '-', color="blue")
    ax.xaxis.set_major_locator(MultipleLocator(5))
    ax.xaxis.set_minor_locator(MultipleLocator(2.5))
    ax.yaxis.set_major_locator(MultipleLocator(0.1))
    ax.yaxis.set_minor_locator(MultipleLocator(0.05))
    plt.xlabel(r'Energy of incident photon (eV)')
    plt.ylabel(r'XAS Intensity (a.u.)')
    plt.text(852, 0.52, r'$L_{3}$', fontsize=20)
    plt.text(869.5, 0.15, r'$L_{2}$', fontsize=20)
    plt.text(846.5, 0.55, r'(a)', fontsize=25)
    plt.subplots_adjust(left=0.15, right=0.95, bottom=0.13, 
                        top=0.95,  wspace=0.05, hspace=0.00)
    plt.savefig('xas_ni.pdf')

def rixs(eval_i, eval_n, T_abs, T_emi):
    gs=list(range(0, 3))
    prob = boltz_dist([eval_i[i] for i in gs], 300) 
    off=857.4
    phi, thin, thout = 0, 15/180.0*np.pi, 75/180.0*np.pi
    # Life-time broadening of the intermediate 
    # (with a core-hole) and final states, respectively 
    Gam_c, Gam = 0.2, 0.1 
    pol = [(0,0), (0,np.pi/2.0)] # pi-pi and pi-sigma polarization
    omega = np.linspace(-5.9, -0.9, 100)
    eloss = np.linspace(-0.5, 5.0, 1000)
    rixs=np.zeros((len(pol),len(eloss),len(omega)),dtype=np.float)

    # Calculate RIXS
    for i, om in enumerate(omega):
        F_fi=scattering_mat(eval_i,eval_n,T_abs[:,:,gs],T_emi,
                           om,Gam_c)
        for j, (alpha, beta) in enumerate(pol):  
            ei, ef=dipole_polvec_rixs(thin,thout,phi,alpha,beta)
            F_mag=np.zeros((len(eval_i),len(gs)),dtype=np.complex)
            for m in range(3):
                for n in range(3):
                    F_mag[:,:]+=ef[m]*F_fi[m,n]*ei[n]
            for m in gs:
                for n in range(len(eval_i)):
                    rixs[j,:,i]+=(prob[m]*np.abs(F_mag[n,m])**2 
                    *Gam/np.pi/((eloss-(eval_i[n]-eval_i[m]))**2 
                                 +Gam**2 )) 

    # plot RIXS map
    plt.figure()
    ax = plt.subplot(1,1,1)
    a,b,c,d=min(omega)+off,max(omega)+off, min(eloss), max(eloss)
    plt.imshow(rixs[0]+rixs[1],extent=[a,b,c,d],
               origin='lower', aspect='auto',cmap='rainbow', 
               interpolation='bicubic')
    ax.xaxis.set_major_locator(MultipleLocator(1))
    ax.xaxis.set_minor_locator(MultipleLocator(0.5))
    ax.yaxis.set_major_locator(MultipleLocator(1))
    ax.yaxis.set_minor_locator(MultipleLocator(0.5))
    plt.xlabel(r'Energy of incident photon (eV)')
    plt.ylabel(r'Energy loss (eV)')
    plt.text(851.6, 4.5, r'(b)', fontsize=25)
    plt.subplots_adjust(left=0.1, right=0.95, bottom=0.13, 
                      top=0.95,  wspace=0.05, hspace=0.00)
    plt.savefig("rixs_map_ni.pdf")

if __name__ == "__main__":
    print("edrixs >>> ED ...")
    eval_i, eval_n, T_abs, T_emi = ed()
    print("edrixs >>> Done ! ")

    print("edrixs >>> Calculate XAS ...")
    xas(eval_i, eval_n, T_abs)
    print("edrixs >>> Done ! ")

    print("edrixs >>> Calculate RIXS ...")
    rixs(eval_i, eval_n, T_abs, T_emi)
    print("edrixs >>> Done ! ")
\end{lstlisting}

\begin{figure}[!ht]
\includegraphics[width=0.48\textwidth]{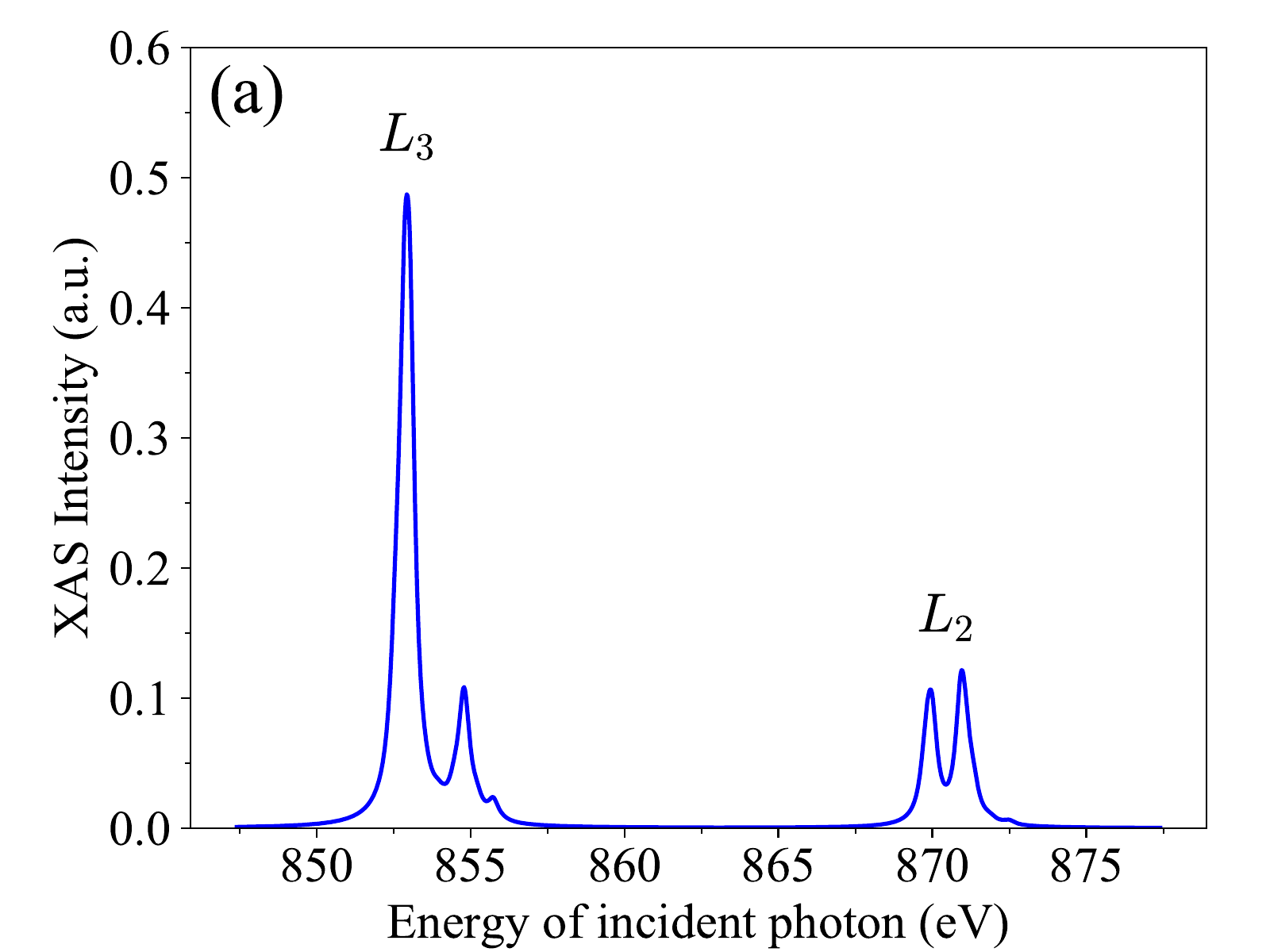}
\includegraphics[width=0.48\textwidth]{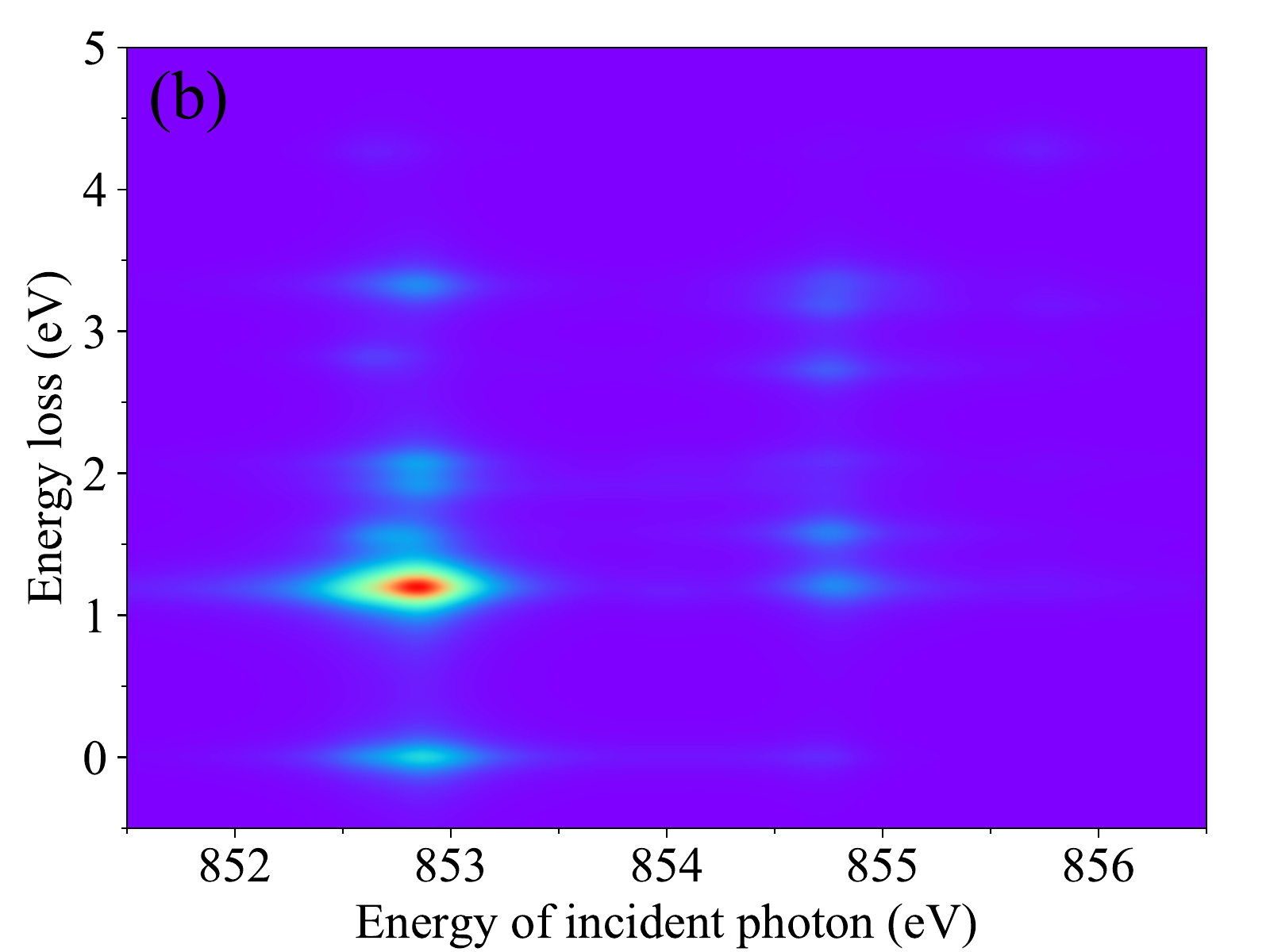}
\caption{(a) XAS spectrum of Ni (b) a RIXS map at $L_{3}$ edge.}
\label{fig:ex1}
\end{figure}

The calculated XAS spectrum and a RIXS map at $L_{3}$ edge are plotted in Fig.~\ref{fig:ex1}.
More results can be found in Ref.~\cite{fabbris:2016}.

\subsection{Two-site Ir-Ir cluster: dimer excitations\label{sec:dimer}}
\begin{figure}[!ht]
\centering
\includegraphics[width=0.5\textwidth]{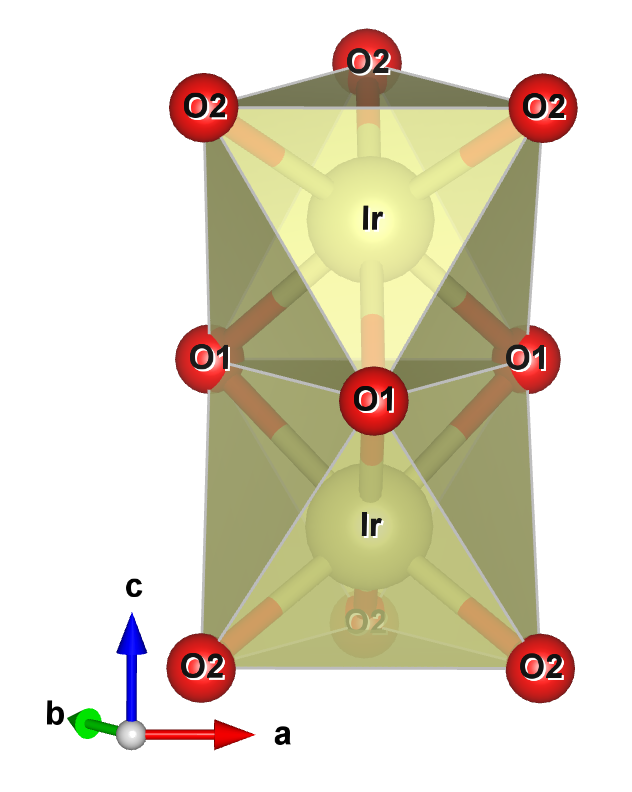}
\caption{The illustration of the face sharing of two IrO$_6$ octahedras.}
\label{fig:face}
\end{figure}

In this section, we show an example of two-sites Ir-Ir cluster model, where two IrO$_6$ octahedras share their face, see Fig.~\ref{fig:face}.
This type of structure can be found in materials, such as Ba$_5$AlIr$_2$O$_{11}$~\cite{Terzic2015coexisting,Streltsov_prb_2017_DFT_BAIO} 
and the 6H-hexagonal oxides Ba$_3AB_2$O$_9$ ($A$=In, Y, Lu, Na and $B$=Ru, Ir)~\cite{Ba3InIr2O9_2017,Ba3MRu2O9_2017,Ba3NaRu2O9_2012}.
Such face sharing structure leads to strong hopping between these two Ir sites, and dimer excitations have been observed in experimental RIXS spectra and confirmed
by a two-site model simulation~\cite{dimer_excitation:2018}. 
Here, we show how to do the calculations step by step. We will simulate RIXS spectra at Ir-$L_{3}$ edge for the cases with and without hopping
between these two Ir sites based on a $t_{2g}$ two-sites model. 
More details of the methods and the Hamiltonian of the on-site Coulomb interaction, spin-orbit coupling (SOC), trigonal crystal field and the hoppings between the two Ir sites
can be found in Ref.~\cite{dimer_excitation:2018}. 

We first go to the \textit{examples} directory and run the Python script,
\begin{verbatim}
    $ cd ${EDRIXS_DIR}/examples/cpc/two_site_cluster
    $ ./get_inputs.py
\end{verbatim}
to do the following things,
\begin{enumerate}
\item It first makes four directories: \textit{ed} for performing ED calculation, \textit{xas} for performing XAS calculation, \textit{rixs\_pp} for performing RIXS calculation 
with $\pi$-$\pi$ polarization and \textit{rixs\_ps} for performing RIXS calculation with $\pi$-$\sigma$ polarization. 
\item Set the control parameters for \textit{ed.x}, \textit{xas.x} and \textit{rixs.x} in file \textit{config.in},
\begin{verbatim}
# File: config.in
&control
num_val_orbs=12   ! Number of total valence orbitals (t2g): 2x6=12
num_core_orbs=12  ! Number of total core orbitals (2p): 2x6=12
ed_solver=0       ! Type of ED solver, full diagonalization of H_{i}
neval=220         ! Number of eigenvalues obtained
nvector=2         ! Number of eigenvectors obtained
idump=.true.      ! Dump eigenvectors to file eigvec.xxx
num_gs=2          ! Numer of gound states used for XAS and RIXS 
                  ! calculations
linsys_tol=1E-10  ! Tolerance for the termination of solving 
                  ! linear equations
nkryl=500         ! Maximum number of Krylov vectors when building 
                  ! Krylov space
gamma_in=2.5      ! Core-hole life-time in eV
omega_in=-540.4   ! Incident x-ray energy at which RIXS
                  ! calculations are performed
&end
\end{verbatim}
\item Get parameters of hopping and Coulomb interaction $t_{\alpha,\beta}$ and $U_{\alpha,\beta,\gamma,\delta}$, and write them in files: \textit{hopping\_i.in}, 
\textit{hopping\_n.in}, \textit{coulomb\_i.in} and \textit{coulomb\_n.in}.
\item Get Fock basis and write to files: \textit{fock\_i.in}, \textit{fock\_n.in} and \textit{fock\_f.in}. Please note that when setting up the Fock basis, we only consider
the valence orbitals because the core orbitals will be explicitly considered within \textit{ed.x}, \textit{xas.x} and \textit{rixs.x}. Here, for $\hat{H}_{i}$, there will be 9 electrons occupying 12 valence orbitals, and for $\hat{H}_{n}$ with a core-hole, there will be 10 electrons occupying 12 valence orbitals.
\item Get transition operators for XAS and RIXS and write to files: \textit{transop\_xas.in}, \textit{transop\_rixs\_i.in} and \textit{transop\_rixs\_f.in}. 
For XAS, we use isotropic polarization. For RIXS, we do calculations for $\pi$-$\pi$ and $\pi$-$\sigma$ polarization. 
The scattering plane is chosen as $ac$-plane. The incident and scattered angles are $\theta_{\text{in}}=\pi/6$ and $\theta_{\text{out}}=\pi/3$, respectively.
\end{enumerate}

\begin{figure}[!ht]
\includegraphics[width=0.49\textwidth]{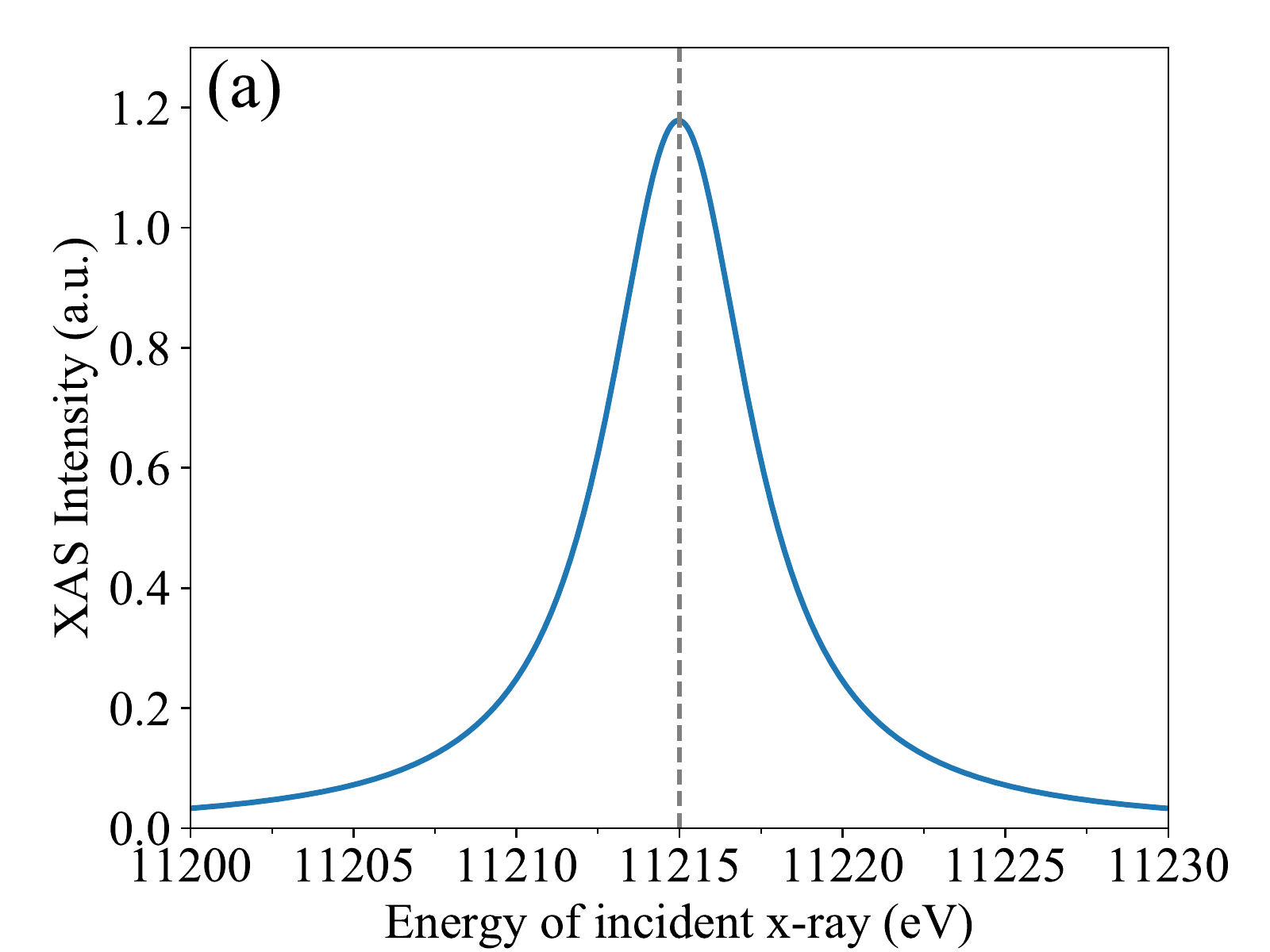}
\includegraphics[width=0.49\textwidth]{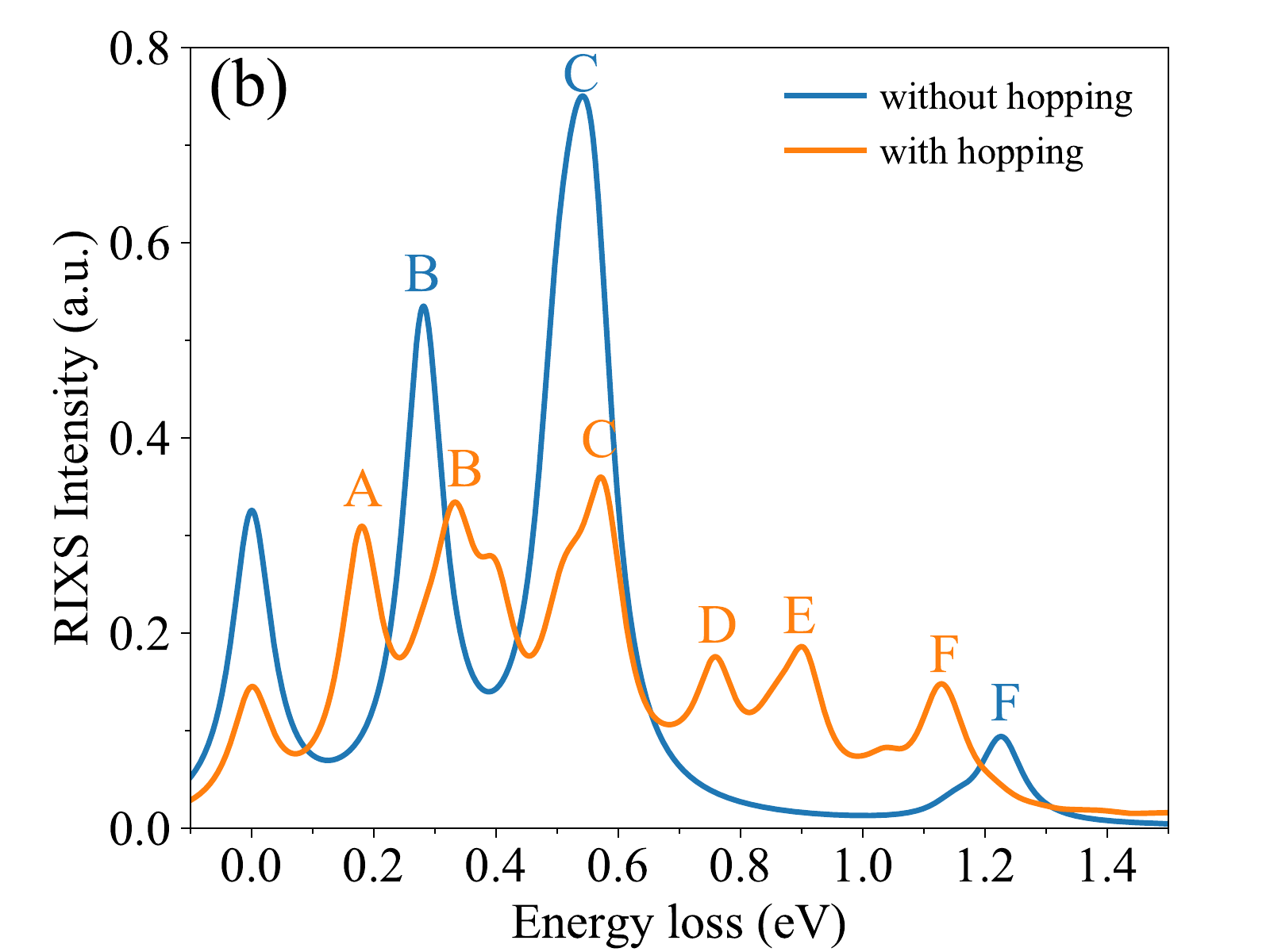}
\caption{(a) Simulated XAS spectrum at Ir-$L_{3}$ edge and (b) RIXS spectrum for the cases with and without hoppings between the two Ir-sites.}
\label{fig:ex2}
\end{figure}

Then, we go to the \textit{ed} directory and run
\begin{verbatim}
    $ mpirun -np 4 ed.x > log.dat
\end{verbatim}
to perform the ED calculation. After ED is finished, we copy eigenvectors to \textit{xas}, \textit{rixs\_pp} and \textit{rixs\_ps} directories,
\begin{verbatim}
    $ cp eigvec.* ../xas
    $ cp eigvec.* ../rixs_pp
    $ cp eigvec.* ../rixs_ps
\end{verbatim}
We can now go to \textit{xas} directory to do XAS calculation and just type
\begin{verbatim}
    $ mpirun -np 4 xas.x > log.dat
\end{verbatim}
After the XAS calculation is finished, we can get the XAS spectrum by running a Python script in upper directory
\begin{verbatim}
    $ get_spectrum.py -N 1000 -ommin -560 -ommax -520 -T 300 -G 2.5  \
    >  -off 11755.5 -f xas_with_hopp.dat xas/*poles*
\end{verbatim}
where, we get the XAS spectrum near Ir-$L_{3}$ edge among energy range [ommin+off,ommax+off].
Temperature is set to be 300 K and $\Gamma_{c}=2.5$ eV.
We can now plot the XAS spectrum, which is shown in Fig.~\ref{fig:ex2}(a).

Finally, we go to the \textit{rixs\_pp} and \textit{rixs\_ps} directories to do RIXS calculations. We need manually change \textit{omega\_in} in file \textit{config.in}
to set the incident x-ray energy at which the RIXS are performed. Here, we set it to be -540.4 eV, which is corresponding to the resonant energy 11.215 keV of the 
Ir-$L_{3}$ edge. We run the command,
\begin{verbatim}
    $ mpirun -np 4 rixs.x > log.dat
\end{verbatim}
to launch the RIXS calculations. After the calculations are finished, we run the Python script 
\begin{verbatim}
    $ get_spectrum.py -N 1000 -ommin -0.2 -ommax 1.5 -T 300 -G 0.04   \
    >  -f rixs_with_hopp.dat rixs_pp/*poles* rixs_ps/*poles*
\end{verbatim}
to get the RIXS spectrum among energy range [-0.2, 1.5]. $\Gamma$ is set to be 0.04 eV.
We repeat the above calculations for the case without hoppings between the two Ir sites by setting $t_{1}=t_{2}=0.0$ in file \textit{get\_inputs.py}.
The RIXS spectrum is plotted in Fig.~\ref{fig:ex2}(b). Without hoppings, we can find three peaks B, C, and F corresponding to the local $d$-$d$ 
excitations which are mainly determined by Hund's coupling $J_{H}$ and SOC $\lambda$. After turning on the hoppings, the peak positions of these $d$-$d$ excitations
will be shifted and new peaks A, D and E corresponding to dimer excitations appear in the RIXS spectrum. The onset of dimer excitations in RIXS spectrum have been
observed in Ba$_5$AlIr$_2$O$_{11}$ experimentally~\cite{dimer_excitation:2018}. Here, we demonstrate that EDRIXS can be helpful to simulate such dimer excitations in materials.

\subsection{Anderson impurity model: charge transfer excitations} 
In this section, we show an example to simulate RIXS spectrum based on Anderson impurity model in an Os compound~\cite{taylor:2017} with nominal $d^{3}$ configuration
to get charge transfer excitations.
We use a $t_{2g}$ Anderson impurity model with three bath sites per orbitals. The bath sites are obtained from a converged DFT+DMFT calculations.
We first go to the \textit{examples} directory and run the Python script,
\begin{verbatim}
    $ cd ${EDRIXS_DIR}/examples/cpc/anderson_impurity
    $ ./get_inputs.py
\end{verbatim}
to build necessary directories, prepare input files just like what we do in Section~\ref{sec:dimer}. 
For Anderson impurity model, we first need to perform ED calculations in subspaces with fixed total occupancy number to 
determine the total occupancy number of the ground states.  
\begin{verbatim}
    $ cd search_gs
    $ echo "mpirun -np 4 ${EDRIXS_DIR}/bin/ed.x" > mpi_cmd.dat
    $ ${EDRIXS_DIR}/bin/search_gs_by_N.py -ntot 24 -nimp 6  \
    > -N1 2 -N2 22 -mpi_cmd mpi_cmd.dat
    $ cd ..
\end{verbatim}
where, the total number of orbitals is 24 and the first 6 are the impurity orbitals. We perform the ED calculations for the total occupancy number ranging from 2 to 22.
For these ED calculations, we only need to get rough results, so we use the ED solver with standard Lanczos algorithm without re-orthogonalization by setting 
\textit{ed\_solver=1} in file \textit{config.in}.

After then ED calculations are done, we find that the total occupancy of the global ground state is 15 in the first line of the file \textit{results.dat}.
Then, we run
\begin{verbatim}
    $ ./get_fock_basis.py -norb 24 -noccu 15
\end{verbatim}
to set up Fock basis.
After that, we can go to the \textit{ed} directory to perform an ED calculation with total occupancy 15. At this time, we use the ED solver with parallel ARPACK library 
to get more accurate ground states by setting \textit{ed\_solver=2}.
\begin{verbatim}
    $ cd ed
    $ mpirun -np 4 ${EDRIXS_DIR}/bin/ed.x > log.dat
    $ cp eigvec.* ../xas
    $ cp eigvec.* ../rixs_pp
    $ cp eigvec.* ../rixs_ps
    $ cd ..
\end{verbatim}

We go to the \textit{xas} directory to perform a XAS calculation and then get the XAS spectrum by running,
\begin{verbatim}
    $ get_spectrum.py  -N 1000 -ommin -20 -ommax 10 -T 50 -G 2.5   \
    >  -off 10877.2 -f xas.dat xas/*poles*
\end{verbatim}
Finally, we can perform RIXS calculations for polarization $\pi$-$\pi$ and $\pi$-$\sigma$.
We first set \textit{omega\_in} to be -6.2, which is corresponding to the resonant energy 10871 eV at Os-$L_{3}$ edge.
\begin{verbatim}
    $ cd rixs_pp
    $ mpirun -np 4 ${EDRIXS_DIR}/bin/rixs.x > log.dat
    $ cd ../rixs_ps
    $ mpirun -np 4 ${EDRIXS_DIR}/bin/rixs.x > log.dat
    $ cd ..
\end{verbatim}
After that, we run
\begin{verbatim}
    $ ${EDRIXS_DIR}/bin/get_spectrum.py -N 1000 -ommin -0.2 -ommax 7   \
    >  -T 50 -G 0.075 -f rixs.dat rixs_pp/*poles* rixs_ps/*poles* 
\end{verbatim}
to get the simulated RIXS spectrum.

\begin{figure}[!ht]
\includegraphics[width=0.49\textwidth]{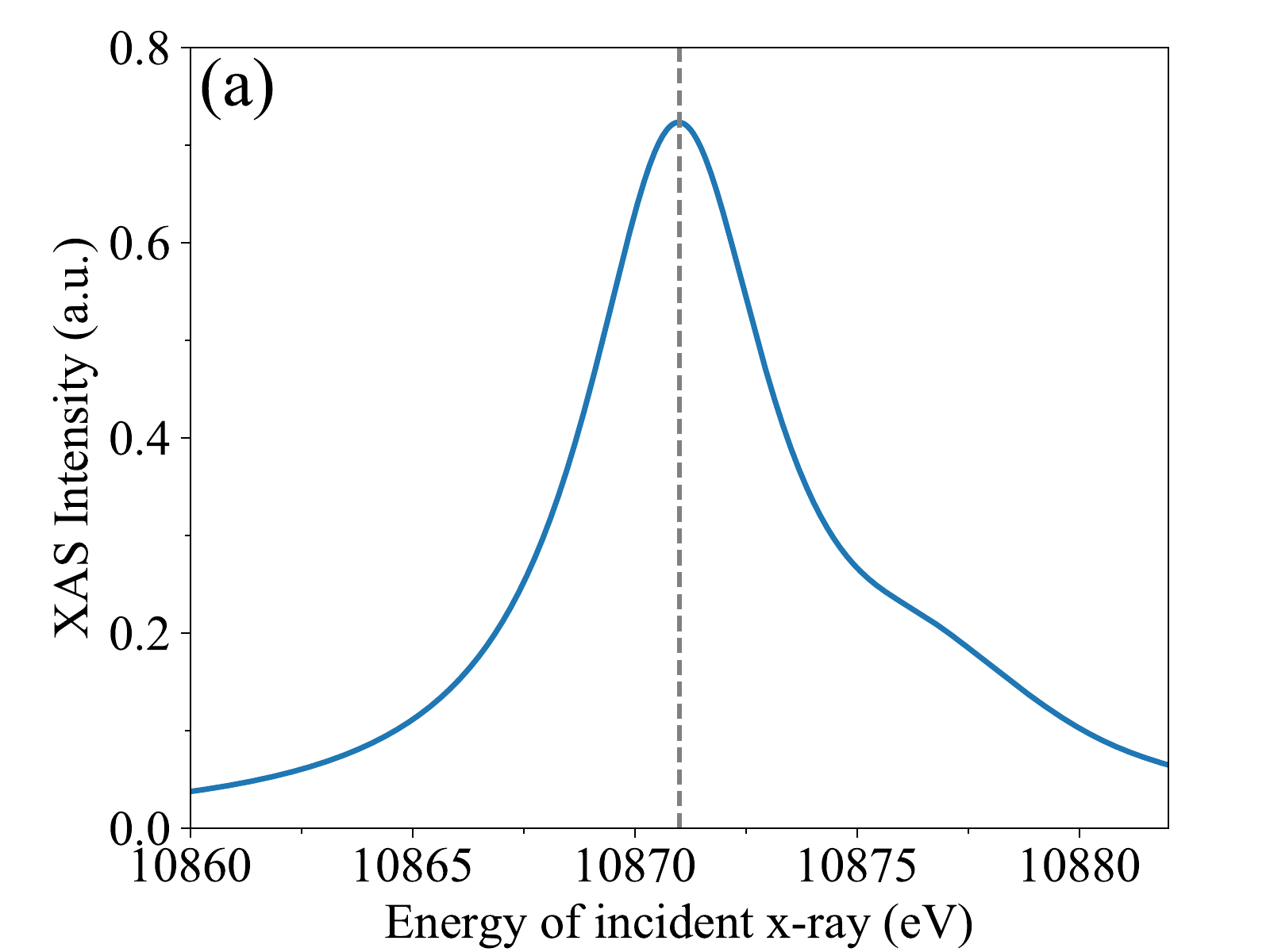}
\includegraphics[width=0.49\textwidth]{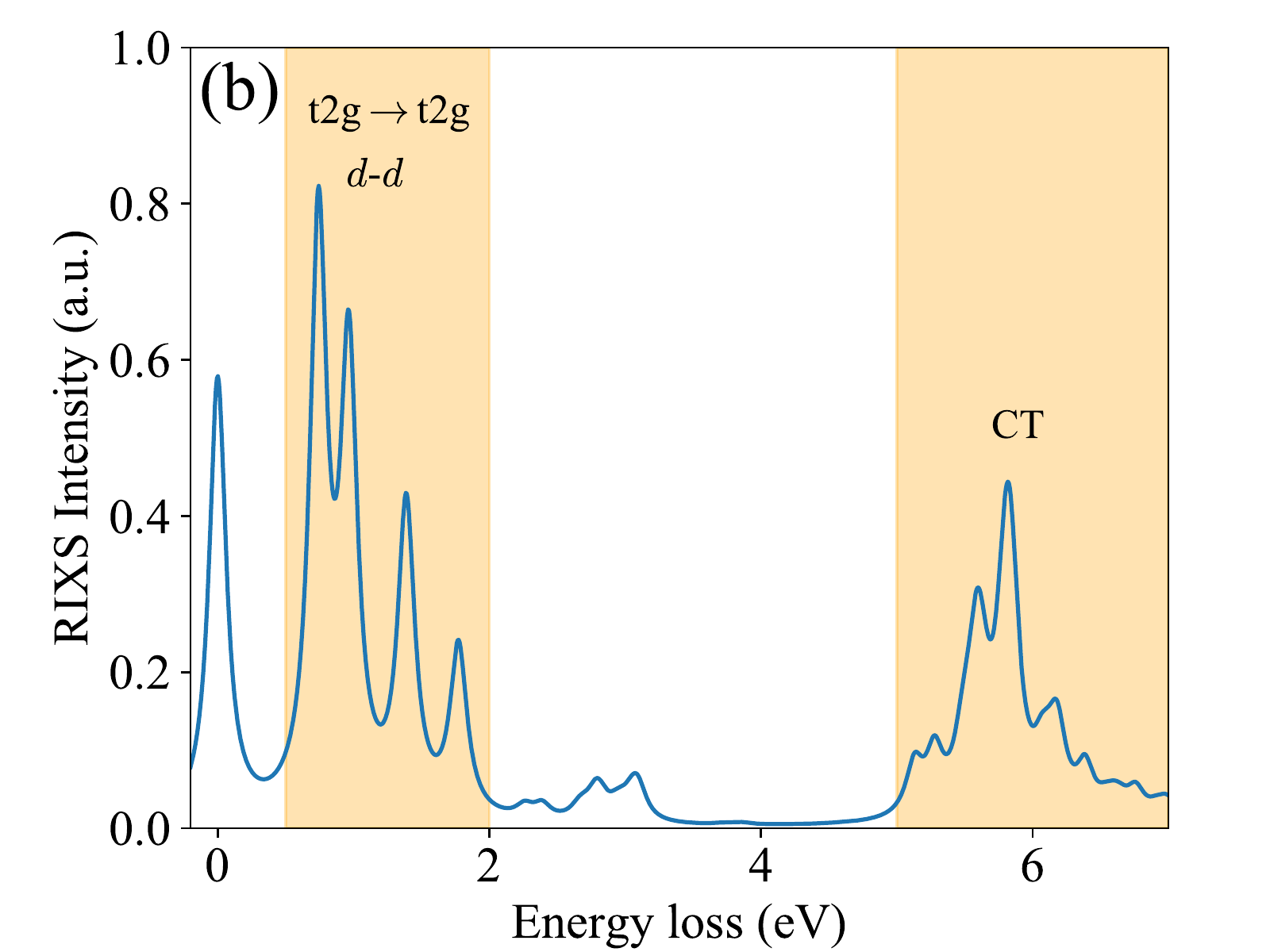}
\caption{(a) Simulated XAS spectrum and (b) RIXS spectrum of an Os compound based on an Anderson impurity model.}
\label{fig:ex3}
\end{figure}

The simulated XAS and RIXS spectrum are plotted in Fig.~\ref{fig:ex3}.
In Fig.~\ref{fig:ex3}(b), we can see four peaks below 2 eV, which are corresponding to $d$-$d$ excitations in the $t_{2g}$ subspace, and a broad charge transfer 
peak in the range 5 $\sim$ 7 eV. More details and results of the RIXS simulations based on DFT+DMFT calculations for this compound will be published in another 
separate paper~\cite{os_dmft_rixs}.

\section{Summary and future developments\label{sec:future}}
In this paper, we introduce the open source toolkit EDRIXS to simulate RIXS spectra. We explain the basic theory and algorithms for RIXS simulations and 
the implementation details of EDRIXS code. We also show three examples to demonstrate its usage. EDRIXS is designed as a platform for theoretical simulations of
x-ray scattering spectroscopy and it will be a very helpful toolkit for the x-ray scattering community.

The development of EDRIXS code is still in progress. The plans of future development are likely to be along the following directions.
More powerful ED solvers based on algorithms such as quantum chemistry (configuration interaction)~\cite{sham:2011, zgid:2012,lin:2013,edsolver:2014}, 
NRG with non-abelian symmetries~\cite{weich:2012} will be implemented to diagonalize even larger size of Anderson impurity model. 
The Python API will be enhanced to provide more powerful and friendly functionalities.
A database for Slater integrals of atoms will be provided. 
Graphical user interface (GUI) can also be implemented for easy use, especially for the experimentalists at beam lines.

\section{Acknowledgments}
We would like to thank very helpful discussions with Hu Miao and Yongxin Yao.
Y.L.W., M.P.M.D. and G.K.\ were supported by the US Department of energy, Office of Science, Basic Energy Sciences as a part of the Computational Materials Science Program through 
the Center for Computational Design of Functional Strongly Correlated Materials and Theoretical Spectroscopy.
G.F. was supported by the U.S. Department of Energy, Office of Basic Energy Sciences, Early Career Award Program under Award No. 1047478. 
Work at Brookhaven National Laboratory was supported by the U.S. Department of Energy, Office of Science, Office of Basic Energy Sciences, under Contract No. DE-SC0012704.

\clearpage

\bibliographystyle{apsrev4-1}
\bibliography{edrixs}

\begin{thebibliography}{63}
\bibitem [{os_()}]{os_dmft_rixs}%
  \BibitemOpen
  \href@noop {} {\bibinfo  {journal} {The manuscript is under preparation}\
  }\BibitemShut {NoStop}%
\end{thebibliography}

\end{document}